\documentclass[fleqn,usenatbib]{mnras}

\usepackage{newtxtext,newtxmath}

\usepackage[T1]{fontenc}
\usepackage{ae,aecompl}
\usepackage{float}
\usepackage{url}
\usepackage{hyperref}
\usepackage{multirow}
\usepackage{subcaption}
\usepackage[para,online,flushleft]{threeparttable}
\usepackage{adjustbox}

\usepackage{graphicx}	
\usepackage{amsmath}	

\def\bw{\Delta\!\nu}
\def\tint{t_\mathrm{nom}}

\title[Observing strategy optimisation]{The Thousand-Pulsar-Array programme on MeerKAT II: observing strategy for pulsar monitoring with subarrays}

\author[Song et al.]{
X.~Song,$^{1}$\thanks{E-mail: xsong@pulsarastronomy.net (XS)}
P.~Weltevrede,$^{1}$
M.~J.~Keith,$^{1}$
S.~Johnston,$^{2}$ A.~Karastergiou,$^{3,4,5}$\newauthor
M.~Bailes,$^{6,7}$
E.~D.~Barr,$^{8}$
S.~Buchner,$^{9}$
M.~Geyer,$^{9}$
B.~V.~Hugo,$^{5,9}$
A.~Jameson,$^{6,7}$\newauthor
A.~Parthasarathy,$^{8}$
D.~J.~Reardon,$^{6,7}$
M.~Serylak,$^{9}$
R.~M.~Shannon,$^{6,7}$
R.~Spiewak,$^{6,7,1}$\newauthor
W.~van Straten$^{10}$ and
V.~Venkatraman~Krishnan$^{8}$
\\
$^{1}$Jodrell Bank Centre for Astrophysics, Department of Physics and Astronomy, University of Manchester, Manchester M13 9PL, UK\\
$^{2}$CSIRO Astronomy and Space Science, Australia Telescope National Facility, PO Box 76, Epping NSW 1710, Australia\\
$^{3}$Department of Astrophysics, University of Oxford, Denys Wilkinson Building, Keble Road, Oxford OX1 3RH, UK\\
$^{4}$Physics Department, University of the Western Cape, Cape Town 7535, South Africa\\
$^{5}$Department of Physics and Electronics, Rhodes University, PO Box 94, Grahamstown 6140, South Africa\\
$^{6}$Centre for Astrophysics and Supercomputing, Swinburne University of Technology, Hawthorn, VIC, 3122 Australia\\
$^{7}$Australia Research Council Centre for Excellence for Gravitational Wave Discovery (OzGrav)\\
$^{8}$Max-Plank-Institut f\"{u}r Radioastronomie, Auf dem H\"{u}gel 69, D-53121 Bonn, Germany\\
$^{9}$South African Radio Astronomy Observatory, 2 Fir Street, Black River Park, Observatory 7925, South Africa \\
$^{10}$Institute for Radio Astronomy \& Space Research, Auckland University of Technology, Private Bag 92006, Auckland 1142, NZ}

\date{Accepted XXX. Received YYY; in original form ZZZ}

\pubyear{2020}

\begin{document}
\label{firstpage}
\pagerange{\pageref{firstpage}--\pageref{lastpage}}
\maketitle

\begin{abstract}

The Thousand Pulsar Array (TPA) project currently monitors about 500 pulsars with the sensitive MeerKAT radio telescope by using subarrays to observe multiple sources simultaneously.
Here we define the adopted observing strategy, which guarantees that each target is observed long enough to obtain a high fidelity pulse profile, thereby reaching a sufficient precision of a simple pulse shape parameter.
This precision is estimated from the contribution of the system noise of the telescope, and the pulse-to-pulse variability of each pulsar, which we quantify under some simplifying assumptions.
We test the assumptions and choice of model parameters using data from the MeerKAT 64-dish array, Lovell and Parkes telescopes.
We demonstrate that the observing times derived from our method produce high fidelity pulse profiles that meet the needs of the TPA in studying pulse shape variability and pulsar timing.
Our method can also be used to compare strategies for observing large numbers of pulsars with telescopes capable of forming multiple subarray configurations.
We find that using two 32-dish MeerKAT subarrays is the most efficient strategy for the TPA project.
We also find that the ability to observe in different array configurations will become increasingly important for large observing programmes using the Square Kilometre Array telescope.

\end{abstract}




\section{Introduction}
Pulsars are stable rotators, their radio emission can be received by radio telescopes as the beam sweeps past the Earth once per rotation period. The radio signal can then be folded by computing a longitude-resolved average of polarised flux to form a pulse profile. The shape of the profile is a unique signature of a pulsar. However, the pulse profile of many pulsars are observed to change between different stable forms, dubbed mode changing \citep{Bartel1982}. This type of rapid change, as well as longer term changes (e.g.\,\citealp{Lyne2010}) can be revealed by pulsar monitoring, provided that the observations are sensitive enough. Information about profile variability is interesting in its own right, and enhances the scientific return of pulsar timing experiments  \citep[e.g.][]{Kramer2006,Lyne2010,Brook2016,Kerr2016,Johnston2020}. 

Pulsar timing experiments, where times of arrival (TOAs) are determined from the pulse profiles, have revealed that pulsars are not perfectly stable rotators. Instabilities are observed in the form of glitches and timing noise. A glitch is a sudden increase in the spin frequency of the pulsar, and is associated with interior processes \citep[see e.g. the review by][]{Haskell2015}. Timing noise is a continuous red noise process in the pulsar spin. 

Timing noise is a common feature in slow (non-recycled) pulsars, and can be revealed with long-term monitoring \citep[see e.g.][]{Hobbs2010,Parthasarathy2019}. At least in some cases timing noise has been associated with long-term pulse profile changes, resulting in a correlation between spin-down rate and profile shape variations \citep[e.g.][]{Lyne2010}. The latter is often quantified by a pulse shape parameter, which measures one particular aspect of the profile shape. These correlations signify global magnetospheric changes, in terms of the amount of current generated in the magnetosphere, and in turn affecting the braking torque \citep[e.g.][]{Kramer2006}. It has been proposed that free precession could be the origin of magnetospheric switching \citep{Jones2012,Kerr2016}. Glitches are also observed to potentially trigger magnetospheric changes, thus affecting the observed pulse profile \citep{Weltevrede2011,Keith2013,Zhao2017}. Identifying timing irregularities and pulse profile variations, and the relation between them, helps understand pulsar interior and magnetospheric processes. Thus, it is important to measure TOAs and pulse shapes representative of the average beam shape of the pulsar to a high precision, where high fidelity pulse profiles are required. In practice, this is often limited by the sensitivity of telescopes, and by the pulse-to-pulse variability of a pulsar.

The statistical noise on a pulse profile can be reduced by using a radio telescope with high sensitivity and wide bandwidth. Care must be taken that the high fidelity profiles obtained are not strongly affected by stochastic pulse shape changes which occur on short timescales, i.e.\ intrinsic variations observed in successive pulses which are unrelated to the long-term magnetospheric state changes. These pulse shape variations lead to an increased scatter in the timing residuals (i.e.\ jitter noise, see \citealt{Shannon2012,Liu2012,Shannon2014}), and can be reduced by averaging more pulses. The amount of jitter noise depends on the pulsar, and can be expected to be higher for pulsars with a large modulation index \citep[see e.g.][]{Weltevrede2006,Burke2012}.

Often, pulsar observations are scheduled in order to achieve a certain signal-to-noise (S/N) of the pulse profile, thereby ignoring the contribution of jitter noise. Requiring a minimum number of pulses helps to compensate for this, although deciding on a specific number can be somewhat ad-hoc. In this paper we propose a scheme to efficiently distribute observing time for a large number of pulsars. Scheduling considerations involving, for example, the sky distribution of pulsars are not addressed here (but see \citealt{Moser2018}). The scheme ensures good sensitivity for detecting pulse shape variations and suitable timing precision. It requires knowledge about parameters of the telescope and the pulsars, and we will demonstrate that basic information of the pulsar readily available from a pulsar catalogue (flux density, period and pulse width) is enough to design an effective observing strategy, as is especially desirable when setting up a monitoring campaign with a new telescope.

This scheme has been created for, and adopted by, the Thousand Pulsar Array (TPA) project on the MeerKAT 64-dish array. This project is a sub-project of MeerTime \citep{Bailes2020}, alongside the Relativistic and Binary Pulsars, Globular Clusters, and the Pulsar Timing Array projects. The TPA project aims to observe more than 1000 non-recycled pulsars to provide high fidelity pulse profiles with full polarisation information. In addition, about 500 pulsars are regularly monitored to study time-dependent phenomena such as glitches, mode-changing, nulling and intermittency. A description of the science goals and early results can be found in Paper~I of the series \citep{Johnston2020}. This paper, Paper~II of the series, describes how the observing times for individual pulsars and the use of subarrays are decided for these two aspects of the observing campaign. Ultimately, the data from a large number of pulsars will allow detailed studies of individual sources, and also help paint a picture of the properties of the pulsar population as a whole. These will be the focus of future publications in the series, which amongst other things will report on measured flux densities, dispersion and rotation measures, scattering properties of the interstellar medium (Oswald et al. submitted), (polarised) profile properties \citep{Serylak2020}, pulse-to-pulse variability and profile state changes.

The goals of the TPA project are made possible by the high sensitivity provided by the 64 dishes of the MeerKAT array \citep{Bailes2020}. This allows the TPA to study profile variability for weaker pulsars, in combination with longer term and higher cadence timing observations from other observatories such as the Lovell Telescope at Jodrell Bank, or the Parkes observatory. With the lower system noise of MeerKAT, the pulse-to-pulse variability dominates the uncertainty in the observed profile for many pulsars, hence compensating for the pulse-to-pulse variability via longer observations is essential. The monitoring of such a large sample of pulsars relies on the possibility to form subarrays, allowing for observation of pulsars in different parts of the sky simultaneously. The proposed scheme provides a framework to assess how one can optimally make use of subarray capabilities to reduce the total observing time. It is expected that this scheme will be especially beneficial for next generation telescopes, such as the Square Kilometre Array (SKA), which aim to observe large numbers of pulsars with different telescope configurations \citep{Smits2009}. 

The structure of the paper is as follows: Section\,\ref{sec:method} explains the proposed time allocation scheme, and in Section\,\ref{sec:ana} the scheme is tested by using data from the MeerKAT 64-dish full array, the Lovell telescope and the Parkes telescope. In Section\,\ref{sec:tpa} the scheme is compared with other observing strategies, and observing strategy for obtaining the TPA legacy dataset is defined. 

\section{Pulse profile fidelity and observing lengths}
\label{sec:method}

We want to obtain high fidelity profiles, so a metric is required to judge the fidelity. Different shape parameters have been used in the literature to quantify pulse profile variations, including the pulse width and the ratio of flux densities of profile components. Here we use the predicted precision of one simple shape parameter to determine the required integration length for a given pulsar. We will demonstrate that this also ensures good precision in the estimation of other shape parameters and pulsar timing. In order to do this, both the system noise and the effect of pulse-to-pulse variability need to be considered.

\subsection{Noise contribution to a single pulse profile bin}
The uncertainty in the flux density of the mean pulse profile in a given longitude bin has two contributions -- one related to the system noise, and one to the pulse-to-pulse variability. The first contribution is determined by the radiometer equation, which predicts the standard deviation (rms) of the system noise distributed across a certain number of bins in the pulse profile to be \citep[e.g.][]{Dicke1946,LK2012}
\begin{equation}
    \sigma_{\mathrm{sys}} = \frac{T_{\mathrm{sys}}}{G \sqrt{n_\mathrm{p}\, \bw \,t}} \sqrt{\frac{n_\mathrm{on} P}{W}}  \,\mathrm{Jy},
    \label{eq:sigma_sys}
\end{equation}
where $G$ is the telescope gain in K/Jy, which depends on the collecting area of the telescope; $T_{\mathrm{sys}}$ is the system temperature in K. $n_\mathrm{p}$ is the number of orthogonal polarisations (usually 2); $\bw$ is the observing bandwidth in Hz; and $t$ is the integration time in seconds. The term $n_{\mathrm{on}}P/W$ quantifies the number of profile bins across a full period $P$ of the pulsar as inferred from the number of independent flux density measurements $n_{\mathrm{on}}$ within the width $W$ of the profile. The system temperature includes the receiver temperature, the cosmic microwave background temperature (2.73 K), atmospheric, ground and sky temperature from Galactic electron synchrotron emission ($T_{\mathrm{sky}}$). At 1.4 GHz, the synchrotron radiation can contribute to more than 20 K near the Galactic centre, but becomes negligible at higher Galactic latitudes. When detected with an optimally matched filter, a periodic signal (i.e. the pulse profile) will have S/N given by
\begin{equation}
\mathrm{S/N}= \frac{S G \sqrt{n_{\mathrm{p}} t \bw}}{T_{\mathrm{sys}}} \sqrt{\frac{P-W}{W}},
\label{eq:snr}
\end{equation}
where $S$ is the mean (pulse period averaged) flux density of the source in Jy.

A large S/N does not necessarily imply a high profile fidelity as intrinsic pulse-to-pulse intensity variability also plays a role. The rms of this contribution, $\sigma_{\mathrm{m}}$, can be quantified by the modulation index $m$ of the pulsar, defined as the standard deviation divided by the mean intensity. 
Since $\sigma_{\mathrm{m}}$ is used to quantify the intrinsic modulation, it is important to exclude the effects of $\sigma_{\mathrm{sys}}$ in the measurement. Further details of how this can be achieved can be found in \citet{Weltevrede2006}, and one can use the tools implemented in \textsc{PSRSALSA}\footnote{\url{http://www.jb.man.ac.uk/~pulsar/Resources/psrsalsa.html}} \citep{Weltevrede2016} to make such measurement. The modulation index is pulse longitude dependent, and is typically smaller at the centre of the profile where the S/N is higher, and becomes larger at the edges of the profile \citep[see e.g.][]{Weltevrede2006,Burke2012}.

For a given $m$, the standard deviation of the flux densities in single pulses is $m\bar{I}$. The mean on-pulse flux density $\bar{I}$ can be inferred from
\begin{equation}
\bar{I} = S \frac{P}{W},
\end{equation}
where $W$ is the equivalent width of the pulse profile, which for simplicity will be approximated with the full-width half maximum $W_{50}$, a value readily available for many pulsars\footnote{\url{http://www.atnf.csiro.au/research/pulsar/psrcat/}} \citep{Manchester2005}. This effectively assumes a top-hat pulse profile, and has the advantage that the actual pulse shape does not need to be quantified. In addition, it will be assumed that the pulse-to-pulse variability behaves, to first order, as white noise such that after integrating $t/P$ pulses, the intrinsic flux density fluctuations obey
\begin{equation}
    \sigma_{\mathrm{m}} = m \bar{I} \sqrt{\frac{P}{t}} = m S P^{3/2} W^{-1} t^{-1/2}.
    \label{eq:sigma_m}
\end{equation}
The total uncertainty in the flux density in a pulse longitude bin in a pulse profile is the quadrature sum of $\sigma_{\mathrm{sys}}$ and $\sigma_{\mathrm{m}}$.

\subsection{Shape parameter statistics}
\label{sec:rms_shape}

The following shape parameter will be considered to judge the fidelity of a pulse profile:
\begin{equation}
\Delta_{\mathrm{shape}} = \frac{I_1-I_2}{\bar{I}}.
\label{eq:shape}
\end{equation}
Here $I_1$ and $I_2$ are the flux densities measured in two profile bins of interest, and $\bar{I}$ is the mean on-pulse flux density. The normalisation is used to correct for effects such as flux density variations due to interstellar scintillation. This particular shape parameter is chosen because of the simplicity in the involved statistics, which can be approximated to be Gaussian. It quantifies relative changes in the flux density at different pulse longitudes, and is closely related to a shape parameter which quantifies the ratio of flux densities \citep[see e.g.][]{Keith2013}. When quantifying actual pulse shape variations from data where the flux density of components change relative to each other, the two bins might correspond to the peaks of two profile components. When predicting the achievable profile fidelity using the framework described here for pulsars without a-priori known intrinsic profile variability, the optimal choice of bin number is unknown. Given the approximation of the profile shape to be a top-hat function, the choice of pulse longitude bins is in fact irrelevant. The uncertainty related to bin choice in actual pulsar data is highlighted in Appendix\,\ref{app:shape_plots}.

The uncertainty in $\Delta_{\mathrm{shape}}$ is quantified as the quadrature sum of the uncertainties of the two bins,
\begin{equation}
\sigma_{\mathrm{shape}} = \frac{ \sqrt{2\left(\sigma_{\mathrm{sys}}^2+\sigma_{\mathrm{m}}^2\right)}}{{\bar I}}, 
\label{eq:on_shapenoise}
\end{equation}
where it is assumed that the variability for the two considered pulse longitude bins is independent, that the modulation index is identical and that the pulse profile shape is a top-hat function. Eq.\,\ref{eq:on_shapenoise} will form the basis on deciding on a suitable integration time for an observation. 

\subsection{Optimal observation lengths}
\label{sec:time_shape}

The optimal observing length is defined to achieve a certain significance of the shape parameter $\Delta_{\mathrm{shape}}$. Combining eqs.\,\ref{eq:sigma_sys} and \ref{eq:sigma_m} in eq.\,\ref{eq:on_shapenoise} gives
\begin{equation}
t = \frac{2 n_\mathrm{on}}{\sigma_{\mathrm{shape}}^2} \frac{W}{S^2P} \left[\frac{(m S P)^2}{n_\mathrm{on} W}+\frac{1}{n_{\mathrm{p}}\bw}\left(\frac{T_{\mathrm{sys}}}{G}\right)^2\right].
\label{eq:time}
\end{equation}

To use eq.\,\ref{eq:time} in practice requires a choice of parameters. The fidelity of the profiles is determined by two parameters: $n_\mathrm{on}$, which defines the resolution at which the profile variability should be resolved and $\sigma_{\mathrm{shape}}$, the ability to measure intensity changes when comparing resolution elements. In what follows, we take $n_\mathrm{on}=16$. This choice implies that profile shape is defined in terms of 16 independent flux density measurements\footnote{This implies that when data are recorded at a higher time resolution (i.e.\ over resolved compared to the resolution $n_{\mathrm{on}}$ at which $\sigma_{\mathrm{shape}}$ is defined), the desired sensitivity in a single bin will only be reached by reducing the time resolution to $n_{\mathrm{on}}$ during post-processing.}. In Section\,\ref{sec:ana} we will demonstrate that taking $\sigma_{\mathrm{shape}}=0.1$, i.e. the ability to detect 10\% changes in intensity at a significance of 1$\sigma$, in combination with this choice of $n_\mathrm{on}$ is reasonable. Although in principle, one could use actual measured values for $m$ for a given pulsar, given that eq.\,\ref{eq:time} relies on other approximations, we demonstrate that taking $m=1$ is reasonable. This choice is motivated by the fact that the modulation index is in general lowest in the most intense parts of the profile, with about 85\% of pulsars having $m$ smaller than 1 in those regions \citep{Weltevrede2006}. Finally, $W$ will be approximated with $W_{50}$, and we use $S_{1400}$ from the pulsar catalogue\footnote{The TPA data will ultimately be used to derive and publish measurements of $S_{1400}$, $W_{50}$ and $m$.}. The sky noise contribution $T_{\mathrm{sky}}$ is also added by extrapolating from the \citet{Haslam1981} all sky map at 408 MHz assuming a spectral index of $-$2.55 \citep{Lawson1987}. 

Hereafter we will refer to the integration time derived from eq.\,\ref{eq:time} using these parameters as $\tint$, and it is explicitly defined as
\begin{equation}
    \tint = 200P+\frac{1600}{\bw}\frac{W_{50}}{P}\left(\frac{T_{\mathrm{sys}}+T_{\mathrm{sky}}}{G S}\right)^2.
    \label{eq:tint}
\end{equation}
This shows that at least 200 rotations of the pulsar are required to ensure $\sigma_{\mathrm{shape}}=0.1$. Since $\tint\propto\sigma_{\mathrm{shape}}^{-2}$, this expression can easily be scaled if a different profile fidelity is required.

Eq.\,\ref{eq:tint} (or eq.\,\ref{eq:time}) can be applied by using parameters which are readily available, but we caution that a large number of simplifying assumptions are made. The pulse-to-pulse variability is taken to be Gaussian distributed, uncorrelated between different pulse longitude bins and for successive pulses. The pulse shape is approximated to be a top-hat function, with a modulation index which is independent of pulse longitude. In addition, this derivation assumes one particular criterion of achieved fidelity of the pulse profile, related to the choice in shape parameter considered. Therefore the accuracy of eqs.\,\ref{eq:time} and \ref{eq:tint} needs to be assessed experimentally, as presented in Section\,\ref{sec:ana}.

\subsection{Exploiting subarrays}
\label{sec:subarray}

The fidelity of profiles is affected both by system noise (related to the second term in brackets in eq.\,\ref{eq:time}) and pulse-to-pulse variability (first term), which both reduce with $\sqrt{t}$. If the pulse-to-pulse variability dominates, the profile fidelity will not be significantly worse when using a less sensitive telescope. If one uses an interferometer, it can therefore be beneficial to use subarrays since multiple pulsars can be observed simultaneously. In the limit of pulse-to-pulse variability dominating, using $N$ subarrays can save a factor of $N$ observing time. Where the system noise dominates, using $N$ subarrays will require $N^2$ times longer observations. This is only partially compensated for as $N$ pulsars can be observed simultaneously. Hence the observing time is still $N$ times longer by using subarrays. Eq.\,\ref{eq:time} can be used to assess if subarrays are beneficial taking into account both the effects of pulse-to-pulse variability and the changes in the gain of different choices of subarrays. 
Other considerations might also affect the choice of the number of subarrays to use. For example, when studying pulse-to-pulse variability one would like to achieve the highest S/N per individual pulse, for which the full array is most suitable.

\section{Verification of the method}
\label{sec:ana}
In order to evaluate the effectiveness of the choice of $\tint$ in eq.\,\ref{eq:tint}, the profile fidelity is tested by observations obtained from the TPA project using the MeerKAT full array, as well as from legacy datasets of the Lovell Telescope at Jodrell Bank Observatory and the Parkes radio telescope. The assumed system parameters for the three telescopes are given in Table\,\ref{tab:para}. Parameters for the pulsars, including the pulse period $P$, the flux density $S_{1400}$ and the full-width half-maximum of the profile $W_{50}$ are taken from the ATNF catalogue (version 1.61, \citealp{Manchester2005}).

\begin{table}
\caption{Telescope parameters used to estimate the system noise. The telescopes considered are the full array of MeerKAT (MeerKAT64, with up to 64 dishes), Lovell and Parkes. For the MeerKAT telescope, the parameters are also given for subarrays with half the number of dishes (MeerKAT32, with up to 32 dishes), and a quarter of the number of dishes (MeerKAT16, with up to 16 dishes).}
\begin{adjustbox}{max width=\linewidth}
\begin{threeparttable}
\begin{tabular}{ccccccc}
        \hline \hline
        Parameter & MeerKAT64 & MeerKAT32 & MeerKAT16 & Lovell & Parkes\\ \hline 
     $G$ (K/Jy) & 2.4 & 1.2 & 0.6 & 1\tnote{a} & 0.69 \\ 
     $T_{\mathrm{sys}}$ (K) & 18 & 18 & 18 & 30\tnote{a} & 25 \\ 
     $\bw$ (MHz) & 775/642\tnote{b} & 775/642\tnote{b} & 775/642\tnote{b} & 384 (200)\tnote{c} & 256 (200)\tnote{c} \\ \hline
\end{tabular}
\begin{tablenotes}
    \item[a] The sensitivity of Lovell is strongly affected by elevation. \\
    \item[b] The quoted MeerKAT values are the fold mode bandwidth (recording 8-second subintegration directly, before the slash), and the single-pulse mode bandwidth (after the slash). \\
    \item[c] The values in parenthesis are the typical effective bandwidth of the Lovell and Parkes telescope, which will be used in calculations.
\end{tablenotes}
\end{threeparttable}
\label{tab:para}
\end{adjustbox}
\end{table}

\subsection{Observations}
\label{sec:obs}

\subsubsection{The MeerKAT telescope}

The MeerKAT gain and system temperature are adopted from \citet{Johnston2020}. Since the number of operational dishes varies, the assumed gain is lower than the maximum sensitivity of the array (see also the MeerKAT specifications\footnote{\url{https://skaafrica.atlassian.net/servicedesk/customer/portal/1/topic/bc9d6ad2-8321-4e13-a97a-d19d6d019a1c/article/277315585}}). The MeerKAT telescope can form subarrays, and two subarray modes are considered here: two subarrays containing half the number of available dishes, and four containing a quarter of the number of dishes (defined as MeerKAT32 and MeerKAT16 in Table\,\ref{tab:para}). The gain is assumed to be half and a quarter of the gain of the full array respectively. These arrays will be referred to as the 64-, 32- and 16-dish arrays, despite that the actual number of involved dishes can vary slightly. 

The MeerKAT data are recorded in fold mode and single-pulse mode simultaneously, and are processed using offline pipelines. Radio frequency interference (RFI) affected frequency channels were discarded and de-dispersion was applied. The data were centred at 1284 MHz with 775 and 642 MHz bandwidth for the fold mode and single-pulse mode, respectively. The corresponding bandwidth of the used MeerKAT data is adopted when computing $\tint$. The effective bandwidth of MeerKAT is very close to the full bandwidth, so the full bandwidth is assumed in calculations.

\subsubsection{The Lovell telescope}
The Lovell observations spanned from 2011 to 2020, recorded using the digital filterbank backend (DFB). The observations were centred at 1520 MHz, with a bandwidth of 384 MHz. Subintegrations of about 10 seconds duration were formed. Dedispersion and RFI mitigation were applied offline. The typical effective bandwidth as quoted in the table are used in calculating $\tint$.

\subsubsection{The Parkes telescope}
The  Parkes observations spanned from 2006 to 2020, and were carried out with the multibeam receiver \citep{Staveley1996} at centre frequency of 1369 MHz with 256 MHz bandwidth. The data were recorded using the digital filterbank backend (PDFB1 to PDFB4, see e.g.\,\citealt{Manchester2013}). Subintegrations with 30 seconds duration were formed. Frequency channels badly affected by RFIs were removed. The most affected subintegrations were also flagged and removed. As for the Lovell telescope, the typical unaffected bandwidth is quoted in the table.

\subsection{Distribution of measured signal-to-noise ratio}
\label{sec:snr}
As profile fidelity does not simply correspond to fixed S/N, it is useful to see the distribution of S/N obtained with integration lengths $\tint$. Here we used the TPA fold mode data observed with the 64-dish array before 20th February 2020. Their observing lengths $t_{\mathrm{obs}}$ are typically longer than $\tint$, in order to collect a larger number of single pulses as aimed for in the pulsar census (see Section\,\ref{sec:tpa}). For each pulsar, the longest fold mode observation was selected and the S/N was computed using \textsc{psrstat} from \textsc{PSRCHIVE}\footnote{\url{http://psrchive.sourceforge.net/}} \citep{Hotan2004} with the \textsc{pdmp} method. The S/N estimates were then scaled by $\sqrt{\tint/t_{\mathrm{obs}}}$.

\begin{figure}
    \centering
    \includegraphics[width=\linewidth]{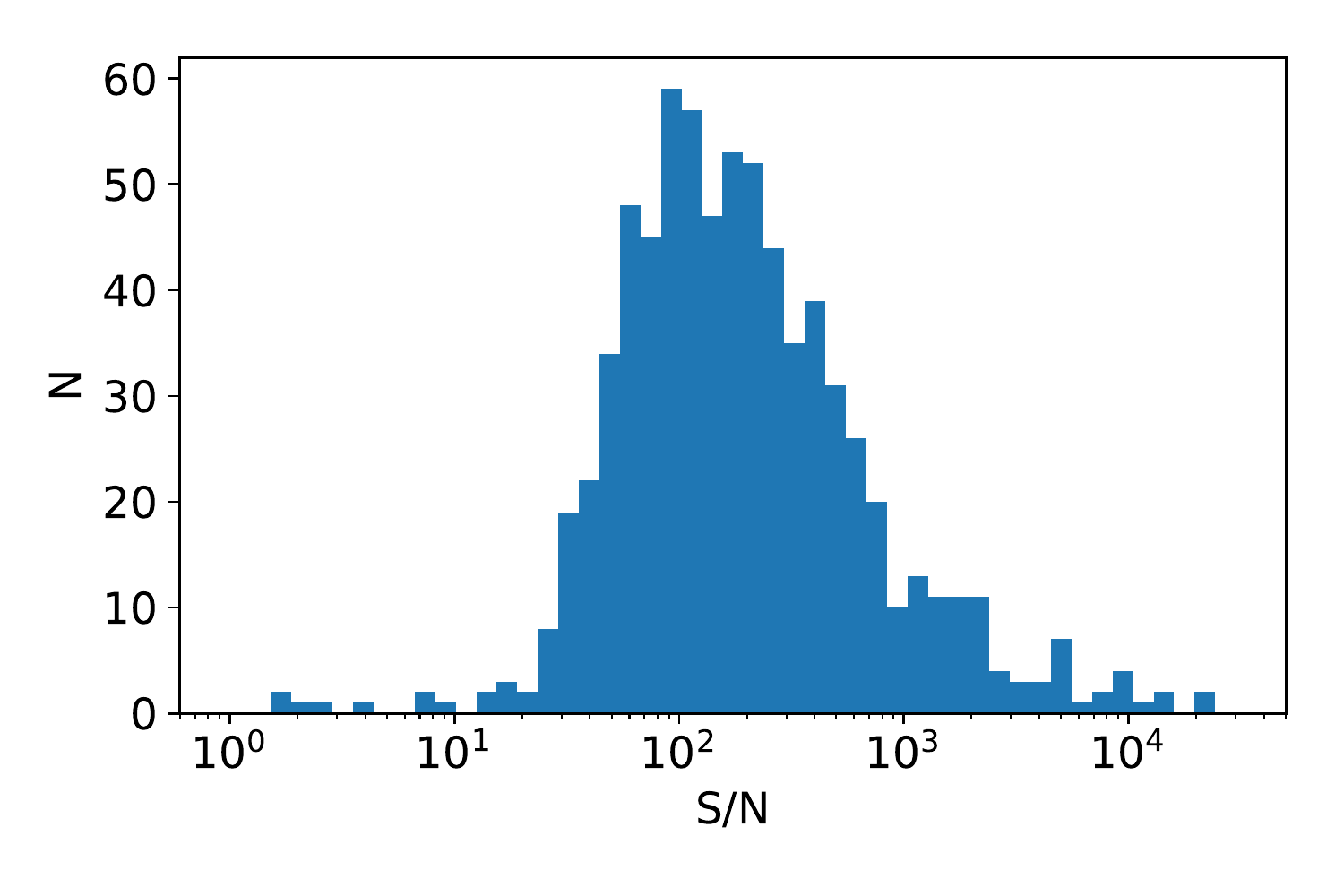}
    \caption{Distribution of S/N estimated for 739 pulsars using the TPA fold mode observations scaled to an observation duration of $\tint$. A bandwidth of 775 MHz is assumed in calculating $\tint$.}
    \label{fig:distsnr}
\end{figure}

The resulting distribution of S/N is shown in Fig.\,\ref{fig:distsnr} and has a median of $\sim$200, with a tail to high values caused by pulsars that are heavily dominated by pulse-to-pulse variability, and with a small number of very low S/N measurements. For the S/N below 10, we find that in all cases these are known nulling or intermittent pulsars for which no reliable detection was made, therefore the flux densities obtained from the catalogue are not representative for the actual flux density during the observation. For most pulsars the uncertainty in the flux density will dominate the uncertainty in the optimum $\tint$ to use (see also \citealp{Levin2013} for discussion of the reliability of catalogue fluxes).

\subsection{Obtained $\Delta_{\mathrm{shape}}$ accuracy}
\label{sec:sigma_shape}
The definition of $\tint$ (eq.\,\ref{eq:tint}) is such that for an observation conducted with this length, a given precision of the shape parameter $\Delta_{\mathrm{shape}}$ (eq.\,\ref{eq:shape}) can be expected under a set of simplifying assumptions. The accuracy of $\Delta_{\mathrm{shape}}$ can be measured directly from observations by quantifying its uncertainty $\sigma_{\mathrm{shape}}$, allowing the accuracy of eq.\,\ref{eq:tint} to be tested directly. 

Here TPA single-pulse data with $t_{\mathrm{obs}} \geq 5 \tint$ (all observed with the full MeerKAT array) were used to measure $\sigma_{\mathrm{shape}}$. This was achieved by extracting blocks of data with length $\tint$ from each dataset, allowing an average profile to be formed for each block. These profiles were normalised independently using the average intensity within an on-pulse region defined by $W_{50}$ as determined from the full observation.

Therefore for each pulsar there were at least five independent profiles accumulated over $\tint$. By measuring the standard deviation of measurements of $\Delta_{\mathrm{shape}}$ gives the uncertainty $\sigma_{\mathrm{shape}}$. The computation of $\Delta_{\mathrm{shape}}$ requires the selection of two on-pulse bins. To assess the accuracy of eq.\,\ref{eq:tint}, we only aim to measure the profile variability related to pulse-to-pulse variability rather than variability related to for example mode changes. Therefore 1000 pairs of bins across $W_{10}$, the pulse width as measured at 10\% of the peak intensity, were selected at random, resulting a distribution of 1000 measurements of $\sigma_{\mathrm{shape}}$ for each observation. 

\begin{figure}
  \centering
  \includegraphics[width=\linewidth]{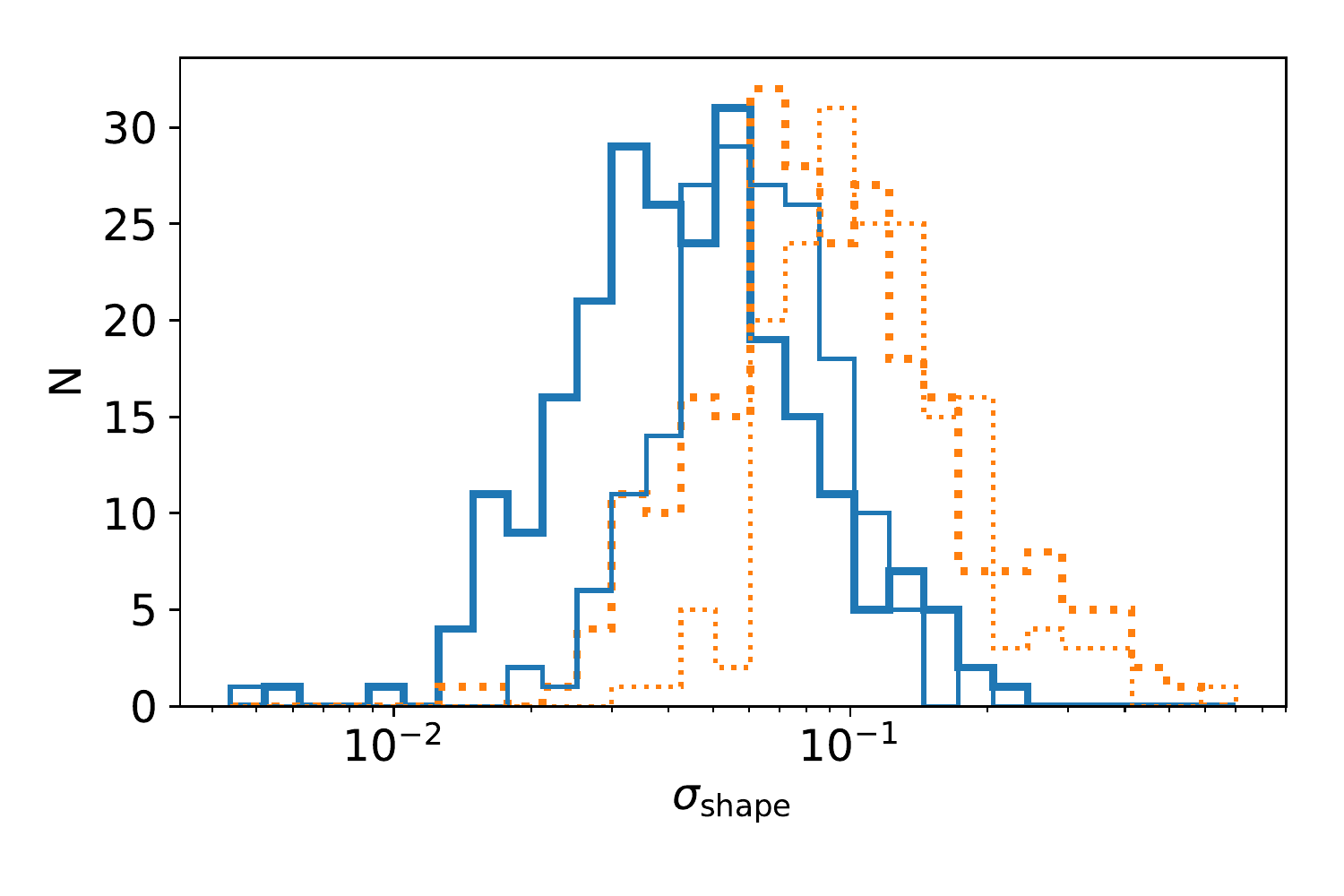}
\caption{Distribution of the measured median (solid thick blue line) and the 95\% percentile (dotted thick orange line) of the $\sigma_{\mathrm{shape}}$ at $\tint$ for 243 pulsars derived from TPA observations using the full array of MeerKAT. The same distribution of the median (sold thin blue line) and 95\% percentile (dotted thin orange line) $\sigma_{\mathrm{shape}}$  for 179 pulsars but with $\tint$ calculated using the measured average modulation index are shown. Four nulling pulsars with large measured values of $\sigma_{\mathrm{shape}}$ were excluded. $\tint$ is determined by assuming a bandwidth of 642 MHz.}
\label{fig:dist_sigma}
\end{figure}

In total 243 pulsars were analysed, and Fig.\,\ref{fig:dist_sigma} shows the distribution of the median (in solid thick blue histogram) and 95\% percentile (in dotted thick orange histogram) of the $\sigma_{\mathrm{shape}}$ distribution obtained for each of them. For the $\tint$ chosen, it is expected that $\sigma_{\mathrm{shape}}=0.1$. One can see that 92 per-cent of pulsars have a median value smaller than 0.1, while 60 per-cent have a 95\% percentile smaller than 0.1. This highlights the fact that $\tint$ is a conservative estimate of the required integration time to achieve a given $\sigma_{\mathrm{shape}}$. After excluding known nulling pulsars, the pulsar with the largest $\sigma_{\mathrm{shape}}$ is PSR~J1913+1011 (median $\sigma_{\mathrm{shape}}\sim 0.22$). Its S/N at $\tint$ is about 30, much lower than the typical S/N of about 200 as seen in Fig.\,\ref{fig:distsnr}, suggesting that the catalogue flux density $S_{1400}$ is overestimated, which results in $\tint$ being underestimated.

The typically small measured $\sigma_{\mathrm{shape}}$ is expected, since as stated in Section\,\ref{sec:time_shape}, many pulsars have a lower modulation index than the assumed $m=1$. To assess the effect of this assumption, the distributions were re-computed by using pulse-phase averaged values of the modulation index as derived from the TPA data\footnote{These values will be reported in a further publication, after the individual datasets have been carefully checked.} using the methodology of \cite{Weltevrede2006}. The resulting distributions of the median $\sigma_{\mathrm{shape}}$ (solid blue) and the 95\% percentile (dotted orange) are shown with thin lines in Fig.\,\ref{fig:dist_sigma}. These distributions are for 179 pulsars, for which $m$ is comfortably detected at least at 10$\sigma$. Indeed, using the measured modulation indices in the calculation of $\tint$ doubles the median of the distribution, bringing it closer to the expected value of 0.1. Furthermore, $\sigma_{\mathrm{shape}}$ as measured will underestimate the actual uncertainty in the obtained profile shape if the pulse-to-pulse variability is correlated between the chosen profile phase bins. It is therefore concluded that using $\tint$ as determined by eq.\,\ref{eq:tint} gives in practice for most pulsars at the very least the expected profile fidelity if $m=1$ is assumed.

We also used observations from the three different telescopes
to verify the scaling of $\sigma_{\mathrm{shape}}$ with telescope gain, modulation index, and integration time as predicted by eq.\,\ref{eq:on_shapenoise}. This analysis is presented in Appendix \ref{app:shape_plots} for four pulsars: PSRs J0729$-$1836 (B0727$-$18), J1803$-$2137 (B1800$-$21), J1829$-$1751 (B1826$-$17) and J1841$-$0345, and the relevant parameters are in Table\,\ref{tab:para_psr}. This analysis shows that the scaling behaves as expected, within the anticipated precision, for these pulsars which cover a large range in the relevant parameter space.

\begin{table}
\caption{Parameters for four selected pulsars. These are the name of the pulsar, pulse period, the ratio of the full-width half-maximum of the profile and the period, flux density, on-pulse weighted modulation index and $\tint$ as calculated using eq.\,\ref{eq:tint} for the MeerKAT full array (MK), Lovell (LT) and Parkes (PKS) telescopes. The bandwidths used are the single-pulse mode bandwidth of 642 MHz for MK, and the effective bandwidths of 200 MHz for LT and PKS.}
\label{tab:para_psr}
\begin{adjustbox}{max width=\linewidth}
\begin{threeparttable}
\begin{tabular}{rcccc}
\hline \hline
Jname & J0729$-$1836 & J1803$-$2137 & J1829$-$1751 & J1841$-$0345 \\
Bname & B0727$-$18 & B1800$-$21 & B1826$-$17 \\
$P$ (s) & 0.51\tnote{a} & 0.13\tnote{b} & 0.31\tnote{c} & 0.20\tnote{d} \\
$W_{50}/P$ & 0.011\tnote{e} & 0.098\tnote{e} & 0.044\tnote{e} & 0.055\tnote{e} \\
$S_{1400}$ (mJy) & 1.9\tnote{f} & 9.6\tnote{g} & 11\tnote{e} & 1.3\tnote{e} \\
$\bar{m}$ & 1.863$\pm$0.003&  0.837$\pm$0.002 & 0.464$\pm$0.001 & 0.629$\pm$0.059\\
$\tint$ MK (s) & 103 & 28 & 62 & 60 \\
$\tint$ LT (s) & 131 & 52 & 67 & 648 \\
$\tint$ PKS (s) & 147 & 69 & 70 & 1066 \\ \hline
\end{tabular}
\begin{tablenotes}
\item[a] \citealt{Hobbs2004};
\item[b] \citealt{Yuan2010}; 
\item[c] \citealt{Hobbs2004}; 
\item[d] \citealt{Morris2002};
\item[e] \citealt{Johnston2018};
\item[f] \citealt{Jankowski2018};
\item[g] \citealt{Rozko2018}.
\end{tablenotes}
\end{threeparttable}
\end{adjustbox}
\end{table}

\subsection{Pulse width measurements}
\label{sec:w50}

While $\tint$ is defined in terms of the expected precision of the shape parameter $\Delta_{\mathrm{shape}}$, a wide range of other metrics can be used to characterise pulse shapes. Nevertheless, many of them can be expected to be measured with roughly the same precision. We illustrate this explicitly by analysing the precision of $W_{50}$ measurements, the full-width at half maximum (FWHM). 

The uncertainty in $W_{50}$, $\sigma_{W}$, was determined by fitting von Mises functions to the observed pulse profiles using \textsc{PSRSALSA}. An analytic model for the profile shape was derived from the full TPA observation in the case of MeerKAT, or a high S/N profile in the case of Lovell and Parkes. Then, available observations were split in equal length pieces, and the model was fit to each of them by allowing only the amplitude and concentration of the von Mises functions to vary. The width was then computed directly from the analytic model, and the standard deviation of the measurements of the individual pieces corresponds to $\sigma_{W}$. 

\begin{figure}
	\centering
	\includegraphics[width=\linewidth]{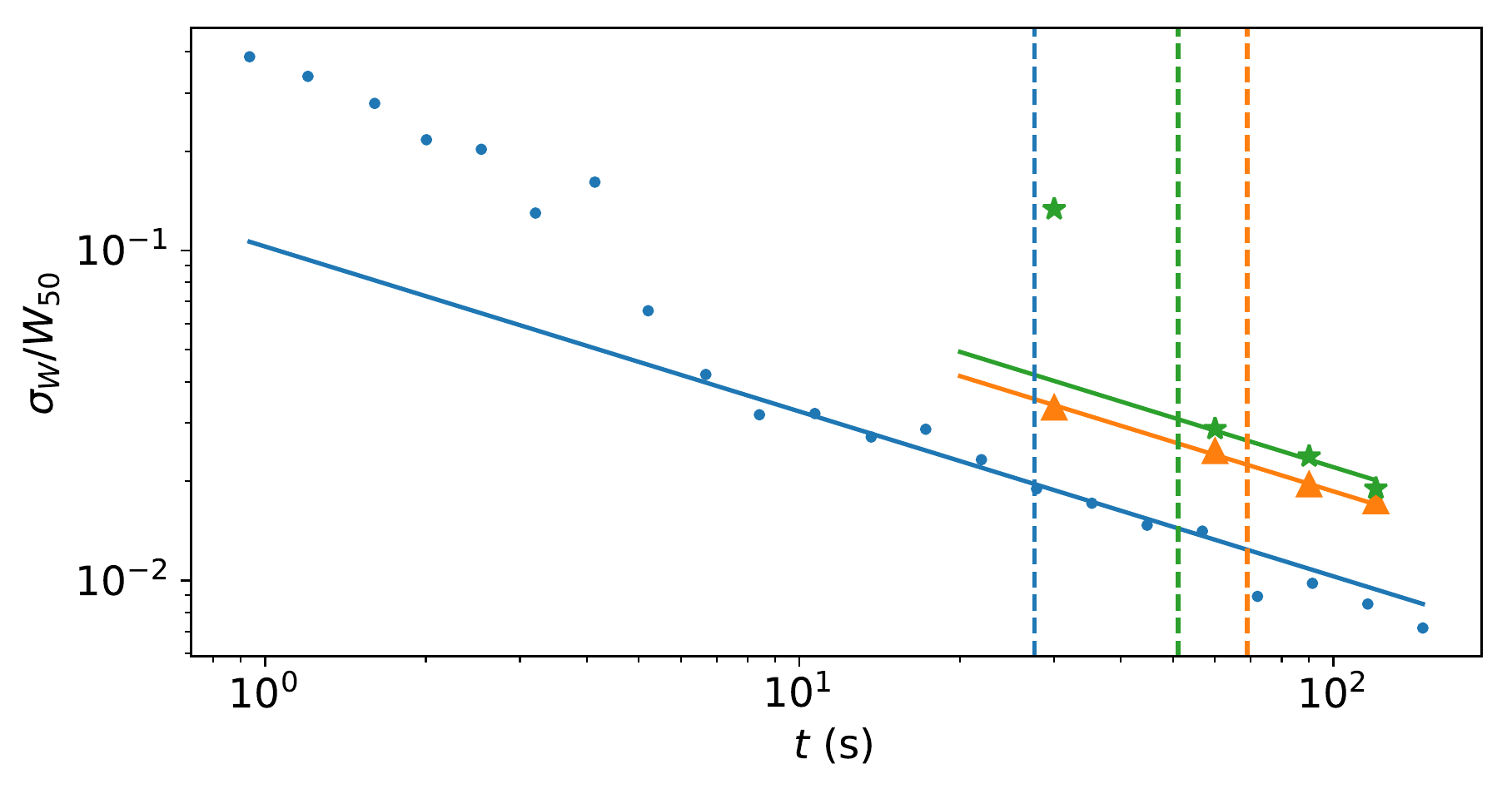}
	\caption{The measured fractional uncertainty of $W_{50}$ for PSR J1803$-$2137 as a function of integration lengths. The results obtained from MeerKAT, Lovell and Parkes data are shown in blue dots, green stars and orange triangles respectively. The diagonal lines show a $1/\sqrt{t}$ dependence to guide the eye. The vertical dashed lines with the same set of colours as the measurements (from left to right corresponding to MeerKAT, Lovell and Parkes telescopes) mark $t=\tint$ (see Table\,\ref{tab:para_psr} for the values).}
	\label{fig:J1803_rmswth}
\end{figure}

Fig.\,\ref{fig:J1803_rmswth} shows the resulting $\sigma_{W}$ for PSR~J1803$-$2137 (normalised by $W_{50}$) as a function of integration time chosen for the individual pieces of data. At long integration times, it can be seen that the precision of $W_{50}$ is the highest for MeerKAT (blue dots), as it has the smallest $\sigma_{W}$ at the same integration time compared to that of the Parkes and Lovell telescope (orange triangles and green stars, respectively). The measured $\sigma_{W}$ is similar for Parkes and Lovell, with the Parkes measurements being slightly more accurate than that of Lovell, while the corresponding $\tint$ is longer (vertical lines). This is because $T_{\mathrm{sys}}$ for the Lovell telescope is higher than assumed for southern hemisphere sources \footnote{For these southern sources $T_{\mathrm{sys}}$ is highly source and observation (elevation) dependent, making it hard to quote a single representative value.}. This is also seen in the $\sigma_{\mathrm{shape}}$ measurements (see Appendix\,\ref{app:shape_plots}). For all three telescopes, the fractional uncertainty follows the $1/\sqrt{t}$ dependence at long enough integration times. For integration lengths well below $\tint$ (as indicated by the vertical dashed lines, see also Table\,\ref{tab:para_psr}), the uncertainty in the width measurements can deviate significantly from this trend. This is because at these short integration times, the pulse shape variability becomes so large that the $W_{50}$ measurements are affected by a different number of profile components in different integrations. For instance, the leading component is sometimes relatively bright, resulting in a large $W_{50}$, while if it is weaker, the measured $W_{50}$ is small as it is determined by the trailing component only. Similarly, for the Lovell data at 30-second integration time (well below $\tint$), the intrinsic profile variability in combination with the relatively high system noise results in the number of components included in the $W_{50}$ measurement to vary. This highlights the complexity involved in the statistics of $W_{50}$ compared to other shape parameters. This is also evident from the analysis for three other pulsars as presented in Appendix\,\ref{app:width}. 

Notwithstanding these complexities, at $\tint$, the fractional uncertainty in $W_{50}$ is somewhat smaller than 0.1. This is also true for the other analysed pulsars. This means that $\tint$ can be expected to give a comparable precision in $W_{50}$ and $\Delta_{\mathrm{shape}}$, hence it demonstrates that the methodology to use $\tint$ to obtain high fidelity profiles is robust. It also shows once more that $\tint$ is chosen somewhat conservatively.

\subsection{Timing precision}
\label{sec:toa}

\begin{figure}
	\centering
	\includegraphics[width=\linewidth]{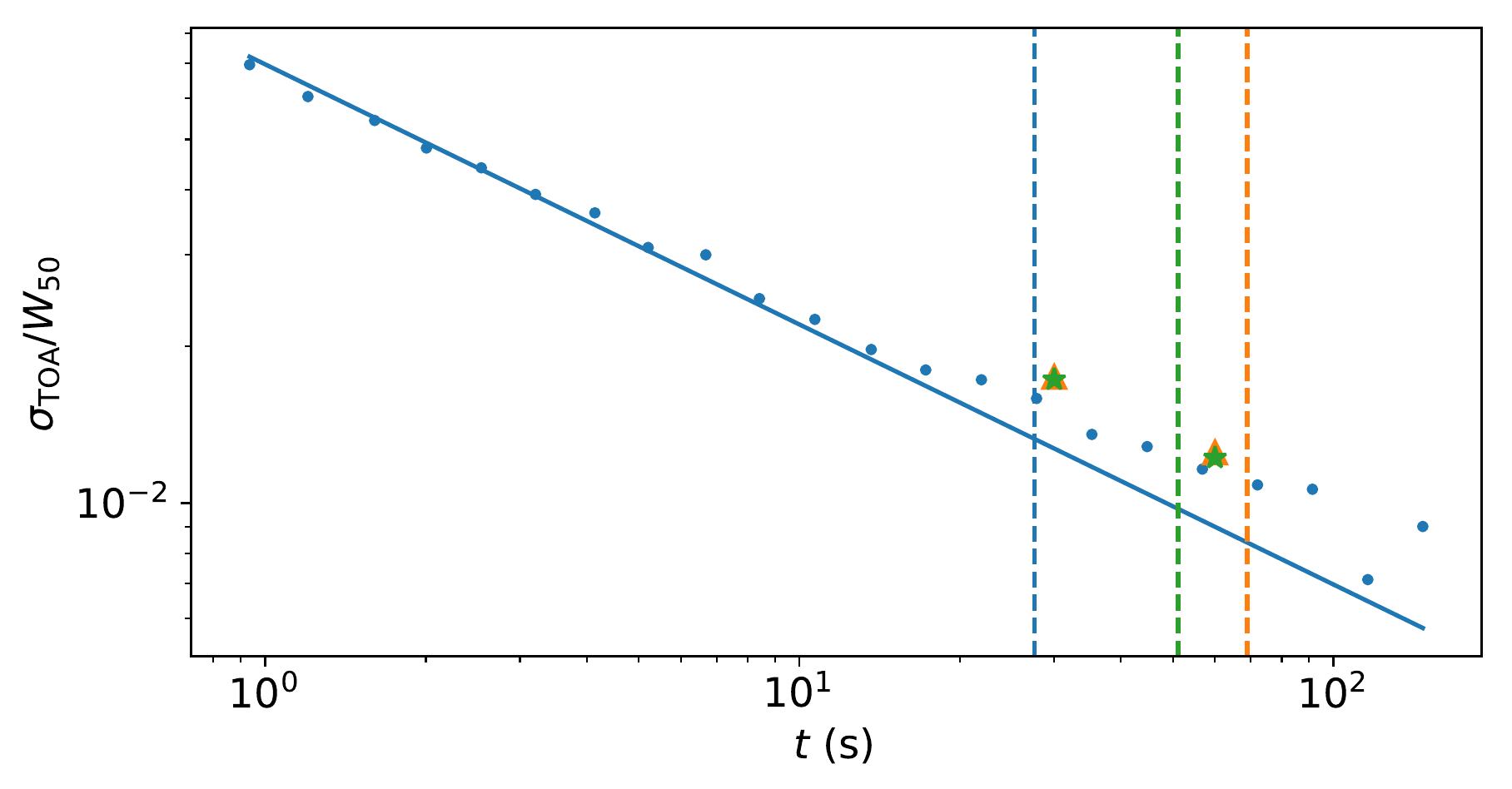}
	\caption{The achieved TOA precision of PSR J1803$-$2137 as a function of integration time, normalised by $W_{50}$. See Fig.\,\ref{fig:J1803_rmswth} for an explanation of the different points and lines, which refer to the MeerKAT, Lovell and Parkes telescopes respectively.}
	\label{fig:J1803_rmsres}
\end{figure}

An observation with duration $\tint$ should ensure that sufficient timing precision can be reached. To test this, we analysed data from all three telescopes for the same four pulsars.
Similar to Section\,\ref{sec:w50}, the data were divided into pieces of a given length $t$. Pulse profiles were produced, and TOAs were determined and compared with timing ephemerides resulting in timing residuals with a standard deviation $\sigma_{\mathrm{TOA}}$ as shown in Fig.\,\ref{fig:J1803_rmsres} for PSR J1803$-$2137. $\sigma_{\mathrm{TOA}}$ quantifies the timing uncertainty arising from profile variability, which is caused by system noise and pulse-to-pulse variability. One can see that the $\sigma_{\mathrm{TOA}}$ is a few per-cent of $W_{50}$ at $\tint$ indicated by the vertical dashed lines (see Table\,\ref{tab:para_psr} for the values).

Despite the gain of MeerKAT being higher than those of the Lovell and Parkes telescopes, the timing precision at a given $t$ are more comparable than for example the precision in $W_{50}$. This is to be expected as the template matching optimally determines the offset of the profile by combining the information of all $n_{\mathrm{on}}$ on-pulse bins. Therefore the timing precision is less affected by white noise compared to shape parameters which attempt to resolve the profile shape, making the high gain of MeerKAT less important for improving the timing precision for pulsars with significant pulse jitter (relatively bright and slowly rotating pulsars). For these pulsars the use of subarrays is beneficial for the TPA (see Section\,\ref{sec:tpa}). Detailed discussions on the effect of pulse-to-pulse variability (jitter noise) in pulsar timing can be found in \citet{Liu2012,Shannon2012,Shannon2014}. The analysis of $\sigma_{\mathrm{TOA}}$ for the other three pulsars is presented in Appendix\,\ref{app:timingrms}. The measured $\sigma_{\mathrm{TOA}}$ are determined to be about or lower than 2 per-cent of $W_{50}$ at $\tint$ for the four pulsars considered for all three telescopes, which is at least as good as the precision achieved for other shape parameters.

\section{Observing strategies: application to the TPA}
\label{sec:tpa}

The TPA project has adopted an observing strategy guided by the method outlined in Section\,\ref{sec:method}. This scheme is contrasted to other possible observing strategies. Some of the conclusions and considerations can be applied to future surveys, including for example a pulsar monitoring survey on the SKA.

The TPA project will produce two legacy datasets, which are defined here for future reference. The first is a census of high fidelity pulse profiles for more than 1000 pulsars. For the brightest 500 pulsars, it is ensured that more than 1000 single pulses are recorded in order to facilitate single-pulse analysis. The second dataset focuses on repeat observations of about 500 bright pulsars for the purpose of pulsar timing and emission variability studies. The strategy for the monitoring campaign will be discussed here.

\begin{figure}
    \centering
    \includegraphics[width=\linewidth]{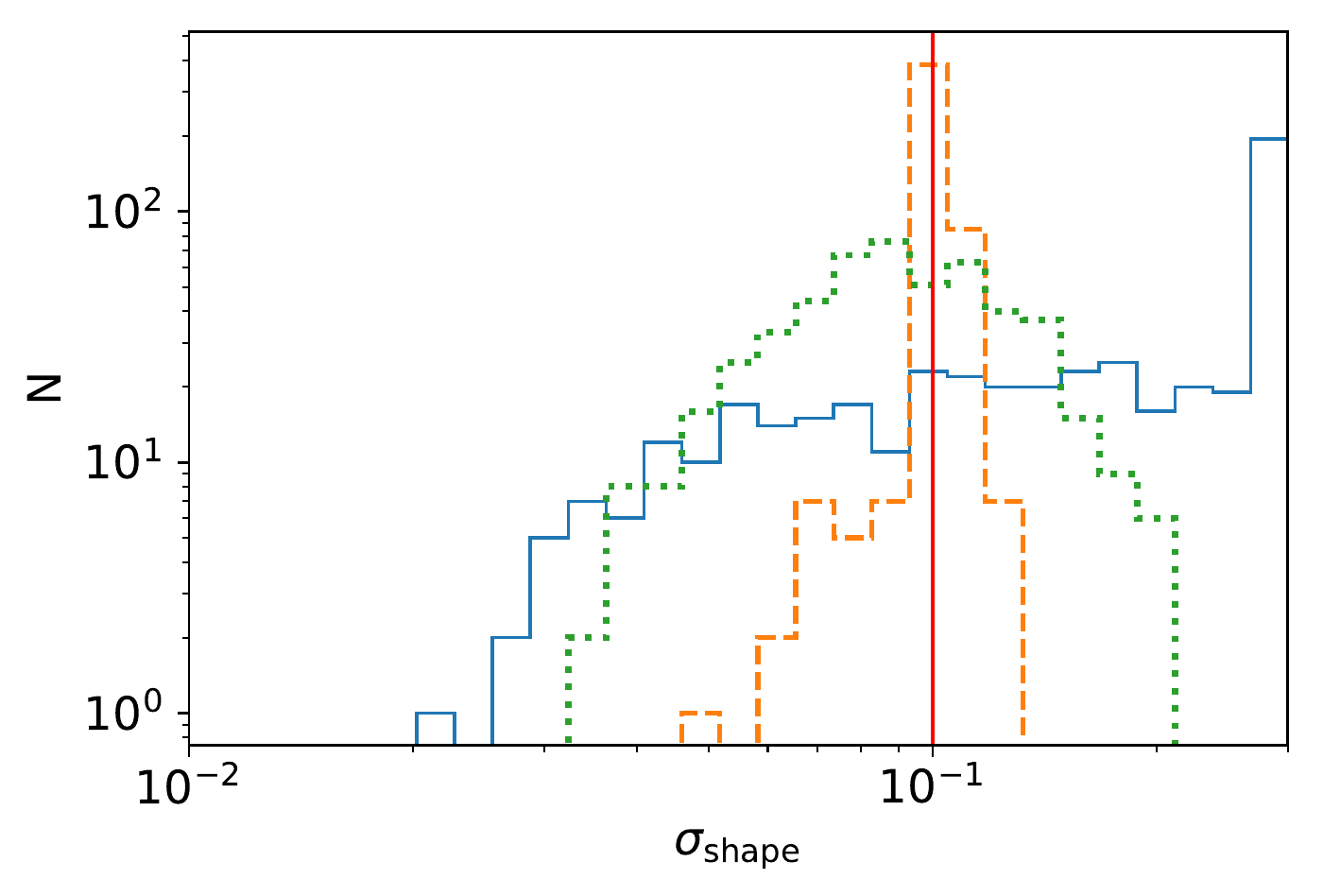}
    \caption{Expected distribution of the uncertainty in the shape parameter $\Delta_{\mathrm{shape}}$ for the brightest 500 TPA pulsars to be observed repeatedly with the MeerKAT full array for different observing strategies with identical total observing time. The adopted TPA observing strategy is defined to have $\sigma_{\mathrm{shape}}=0.1$ for all pulsars (indicated by the red solid line). The blue solid histogram corresponds to a strategy where the S/N is identical (191) for the pulse profiles obtained for each pulsar. This distribution has a long tail which is truncated, resulting in the pile-up at $\sigma_{\mathrm{shape}}=0.3$. The orange dashed histogram corresponds to the strategy where $\mathrm{S/N}=85$, but in addition the integration time should exceed $200P$. The green dotted histogram shows the strategy where a constant integration time ($\sim$134 seconds) is defined for each pulsar.}
    \label{fig:dist_n}
\end{figure}

The adopted TPA strategy is to regularly observe 500 pulsars long enough such that $\sigma_{\mathrm{shape}}=0.1$ is achieved according to eq.\,\ref{eq:tint} (assuming a bandwidth of 775 MHz). To observe these pulsars with the full MeerKAT array would require $\sim$18.7 hours (without overheads). Keeping the total observing time fixed, we can compare this to a strategy that aims to achieve a constant S/N, as defined by eq.\,\ref{eq:snr}. Splitting the observing time between the same 500 pulsars in this way results in a S/N of 191 for each pulsar. As a result, a spread of profile fidelities as defined by $\sigma_{\mathrm{shape}}$ will be obtained. This distribution is shown as the blue solid histogram in Fig.\,\ref{fig:dist_n}. The weakest pulsars are observed longer than $t_{\mathrm{nom}}$ in this scheme, resulting in $\sigma_{\mathrm{shape}}<0.1$. On the other hand, the short durations for the brighter pulsars result in a significant tail of $\sigma_{\mathrm{shape}}>0.1$. For these bright pulsars, the fidelity  will be quite poor despite the high S/N because only the system noise contribution to the pulse profile variability is considered, while the pulse-to-pulse variability is ignored. 

To avoid this problem, one can adopt a minimum required number of pulses to form a stable pulse profile. As predicted by eq.\,\ref{eq:tint}, a minimum of 200 pulsar rotations is expected to be required to achieve $\sigma_{\mathrm{shape}}=0.1$. This value was also adopted by \citet{Smits2009} in their explored SKA monitoring strategy. Taking this value, in combination with setting the obtained S/N to 85, results in the same total observing time of 18.7 hours for the 500 TPA pulsars. This is shown in Fig.\,\ref{fig:dist_n} as the orange dashed distribution. One can see that the spread has been significantly reduced, showing that this strategy is effective in obtaining high fidelity profiles for a large sample of pulsars with a wide range of flux densities. Some of the weaker pulsars are still observed for relatively long, resulting in observations with a $\sigma_{\mathrm{shape}}$ about twice as small as the nominal $\sigma_{\mathrm{shape}}=0.1$.

A very different approach one could take is to observe all pulsars with the same integration length. Such approach is currently taken for example with the Parkes young pulsar timing programme. Such a strategy means that 500 pulsars can be observed for $\sim$134 seconds in 18.7h with the full MeerKAT array. The resulting $\sigma_{\mathrm{shape}}$ distribution is shown as the green dotted histogram in Fig.\,\ref{fig:dist_n}. Taking equal integration times for all pulsars implies that slowly rotating pulsars are observed for a smaller number of periods. The result is that for these pulsars, low-fidelity profiles will be obtained and conversely, for rapidly rotating pulsars excessive profile fidelity is achieve, hence the $\sigma_{\mathrm{shape}}$ distribution is wide.

For the repeat observations within the TPA, there is no requirement for the full array to be used to maximise single pulse sensitivity. Therefore sub-arraying (see Section\,\ref{sec:subarray}) can be exploited to maximise observing efficiency. Observing all 500 pulsars with two 32-dish arrays (gain is reduced by a factor of two, but two pulsars can be observed simultaneously) would reduce the required telescope time to $\sim$11.4 hours (without overheads), compared to $\sim$18.7 hours when the full array is used. More complicated scenarios can be considered. Using a combination of the 32-dish and full array can modestly reduce the required telescope time further to $\sim$11.0 hours. Using a combination of the full 64-dish array, two 32-dish arrays and four 16-dish arrays could reduce this even further to $\sim$7.9 hours. It needs to be stressed that these calculations do not take into consideration the overheads associated with changing array configurations (e.g.\ phasing up of the arrays). Moreover, for sub-arraying to be efficient, each subarray should be allocated observations with similar total duration (for pulsars which are visible at a given time) to allow them to observe simultaneously. This very quickly becomes impractical and inefficient when often switching between array configurations. Therefore, the TPA will use a fixed array configuration of two 32-dish subarrays to monitor the 500 brightest pulsars for legacy dataset 2.

For completeness, the TPA observing strategy for legacy dataset 1, the one-off observations for 1000 pulsars is also defined here. The observations require the use of the full array. This allows the highest fidelity data to be recorded, especially in terms of obtaining the highest S/N of the recorded single pulses. The integration time for the 500 brightest pulsars is set by the requirement that at least 1000 pulses will be recorded, while for the weaker 500 pulsars, the integration time is given by $\tint$. In addition, a minimum length of 90 seconds is applied to make sure that overheads are a relatively modest contribution to the actual observing time. For the weaker short-period pulsars, more than 1000 pulses would be obtained, but not for the longer period pulsars. This dataset would contain in total $\sim$10$^6$ single pulses for the 1000 pulsars. The completion of this campaign is predicted to take $\sim$137 hours of observing time (excluding overheads).

In the future, SKA-mid will consist of $\sim$200 dishes similar to those of MeerKAT. To illustrate how a pulsar monitoring campaign could be designed, we consider a scenario where it is aimed to observe 1000 pulsars (those of TPA legacy dataset 1) in a single observing run\footnote{Of course it is not guaranteed that observing time for such a monitoring campaign will be awarded in large blocks. In addition, the source distribution across the sky plays a role as well and overheads are ignored. So this serves only as an illustration of how one could design such a campaign.} without the need to change array configurations in order to achieve $\sigma_{\mathrm{shape}}=0.1$ for each observation. With MeerKAT this would take 67.2 hours by using the 64-dish array. On the other hand, the larger array provided by SKA-mid will allow more efficient strategies to be devised to save telescope time. By splitting the array into four 50-dish subarrays, the 1000 pulsars can be observed in 21.5 hours. When allowing the subarrays to have unequal sizes (but fixed within the session) the efficiency can be boosted further. For example, by forming one 128-dish subarray and two 36-dish subarrays, the 1000 pulsars can be observed in 18.7 hours of telescope time. Splitting the array in one 110-dish subarray and three 30-dish subarrays, the total telescope time can be reduced further to 18.2 hours. Without splitting the array, observing the same 1000 pulsars would take 40.5 hours. This highlights the important role that subarraying will play in radio astronomy, and we conclude that the employed methodology provides a framework which allows efficient observing strategies to be designed with a minimum set of required pulsar properties to be known a-priori. This framework can also be used to assess the optimal use of different facilities to pursue common science goals.

\section{Conclusion}
\label{sec:discussion}
The efficient design of pulsar monitoring campaigns becomes increasingly important with the construction of ever larger radio telescopes and the ability to exploit subarrays. To aid these efforts, a scheme is developed. This scheme ensures that high fidelity pulse profiles are obtained for useful pulsar timing experiments and to study pulse profile variability.

Our scheme quantifies the required observing time (eq.\,\ref{eq:time}) to obtain a given precision $\sigma_{\mathrm{shape}}$ in the shape parameter $\Delta_{\mathrm{shape}}$ (eq.\,\ref{eq:shape}), under some simplifying conditions. This shape parameter corresponds to the flux density differences between two pulse profile bins. The uncertainty in the shape parameter includes the system noise and pulse-to-pulse variability, and is expressed in terms of telescope and pulsar parameters. This scheme is easy to apply as the parameters used are readily available for many pulsars. Furthermore, the exact profile shape and the pulse phase dependent pulse-to-pulse variability are ignored. The scheme can also help optimally exploit subarrays. For small ratios of the system noise and pulse-to-pulse variability (first and second term in brackets of eq.\,\ref{eq:time}, respectively), subarraying is beneficial. 

For the MeerKAT TPA project, the observing time $\tint$ (eq.\,\ref{eq:tint}, derived from eq.\,\ref{eq:time}) is applied to achieve a high profile fidelity using assumed or chosen parameters. In particular, we take $\sigma_{\mathrm{shape}}=0.1$, the ability of detecting a 10\% change in the shape parameter at 1$\sigma$ significance. In addition, the modulation index $m$, which quantifies the pulse-to-pulse variability, is taken to be 1 for all pulsars. The number of on-pulse bins across $W_{50}$, $n_{\mathrm{on}}=16$ is adopted. These parameters correspond to a typical profile S/N of $\sim$200. From the analysis of 243 pulsars observed with the MeerKAT full array, it is found that the predicted integration time $\tint$ is conservative, i.e.\ typically a higher than required profile fidelity is obtained. This is largely because of the choice of $m=1$ which overestimates the pulse-to-pulse variability of most pulsars.

Data from the MeerKAT full-array, and the Lovell and Parkes radio telescopes were analysed for four pulsars which cover a wide range of the relevant pulsar parameter space. This confirmed the expected scaling with telescope parameters such as the gain. Moreover, although the integration times are related to one particular shape parameter, it is demonstrated for these pulsars that the obtained profiles also give a comparable, or even better precision in pulse width $W_{50}$ and timing residuals (relative to $W_{50}$). Hence the scheme allows to define useful integration times for a large range of science goals.

The science goals of the TPA include the study of profile variability, supported by observations from other observatories such as the Lovell telescope at Jodrell Bank, and the Parkes observatory. For this purpose, the TPA aims to regularly observe 500 bright pulsars. Using these pulsars as an example, different observing strategies are compared with our scheme by determining the expected profile fidelities. It is concluded that the adopted scheme is a suitable way to distribute observing time, and that using two subarrays is the best way to save telescope time. More complex array configurations are deemed inefficient because of the overheads involved in switching array configurations. The use of subarrays in radio astronomy plays an increasingly important role, as is highlighted by considering a scenario where 1000 known pulsars are being monitored with SKA-mid.

\section*{Acknowledgements}
The MeerKAT telescope is operated by the South African Radio Astronomy Observatory, which is a facility of the National Research Foundation, an agency of the Department of Science and Innovation. Pulsar research at Jodrell Bank Centre for Astrophysics and Jodrell Bank Observatory is supported by a consolidated grant from the UK Science and Technology Facilities Council (STFC). The Parkes telescope is part of the Australia Telescope National Facility which is funded by the Commonwealth of Australia for operation as a National Facility managed by CSIRO. XS acknowledges financial support from the President's Doctoral Scholar Award. MeerTime data is housed and processed on the OzSTAR supercomputer at Swinburne University of Technology. MB and RS acknowledge support by the Australian Government through the Australian Research Council (ARC), grants CE170100004 (OzGrav) and FL150100148.  RMS acknowledges support through ARC future fellowship FT 190100155. 

\section*{Data availability}
The data underlying this article will be shared on reasonable request to the corresponding author.

\bibliographystyle{mnras}
\bibliography{tpa}



\appendix

\section{Detailed analysis for four pulsars}

Four pulsars were selected for detailed analysis and their properties are summarised in Table\,\ref{tab:para_psr}, including the computed required integration times $\tint$. These pulsars were selected because suitable MeerKAT, Lovell and Parkes data were available, and because they cover a wide range of pulsar parameters. PSR J0729$-$1836 has the narrowest pulse profile in the sample and its weighted modulation index ($\bar{m}=1.9$) is much larger than what is typically observed for pulsars. Although it is relatively weak, the pulse-to-pulse variability dominates over the system noise for the three considered telescopes. PSR J1803$-$2137 is a relatively bright pulsar with an unusually wide profile.  While its pulse-to-pulse variability is comparable to the system noise for the Lovell and Parkes telescopes, the pulse-to-pulse variability dominates for MeerKAT. PSR J1829$-$1751 is bright and has a narrow pulse profile with $\bar{m}=0.5$. The pulse-to-pulse variability dominates over the system noise for all three telescopes. PSR J1841$-$0345 is the weakest pulsar considered, and the system noise dominates for both the Lovell and Parkes telescopes, while it is comparable to its pulse-to-pulse variability for MeerKAT. 

\subsection{Measured $\sigma_{\mathrm{shape}}$ as a function of integrating length}
\label{app:shape_plots}

In Section\,\ref{sec:sigma_shape}, it is shown that the $\sigma_{\mathrm{shape}}$ obtained from an integration time $\tint$ for a sample of pulsars observed by the MeerKAT 64-dish array are in most cases smaller than specified in the calculation of $\tint$, and it was identified that the assumed modulation index $m=1$ is largely responsible. Here, we compute $\sigma_{\mathrm{shape}}$ as a function of integration time for the four chosen pulsars using data from the three different telescopes. This results in graphs similar to what is shown for $\sigma_{W}$ and $\sigma_{\mathrm{TOA}}$ in Figs.\,\ref{fig:J1803_rmswth} and \ref{fig:J1803_rmsres}, allowing confirmation that the results conform with the expected $1/\sqrt{t}$ scaling, and that the expected precision is achieved at $\tint$.

The method of measuring $\sigma_{\mathrm{shape}}$ as outlined in Section\,\ref{sec:sigma_shape} is repeated for the four pulsars, except that now various integration times are considered. Since the Lovell and Parkes observations span many years, the profiles are firstly aligned by using a template generated from a high S/N pulse profile using \textsc{PSRCHIVE} tool \textsc{paas}. The uncertainty $\sigma_{\mathrm{shape}}$ was determined for 20 pairs of randomly selected profile bins within $W_{10}$. In addition, $\sigma_{\mathrm{off}}$ is determined, which quantifies the contribution of the system noise to $\sigma_{\mathrm{shape}}$. The data were split in $N$ equal length blocks with a given integration time and the off-pulse standard deviation in the profiles were determined for each of them. Gaussian noise with the measured standard deviation was added to the pulse profile obtained from the full observation, resulting in $N$ perturbed profiles. The $\sigma_{\mathrm{shape}}$ was determined as before, which defines $\sigma_{\mathrm{off}}$. As per eq.\,\ref{eq:on_shapenoise}, $\sigma_{\mathrm{off}}=\sqrt{2} \sigma_{\mathrm{sys}}/\bar{I}$.

\begin{figure*}
	\centering
	\includegraphics[width=\linewidth]{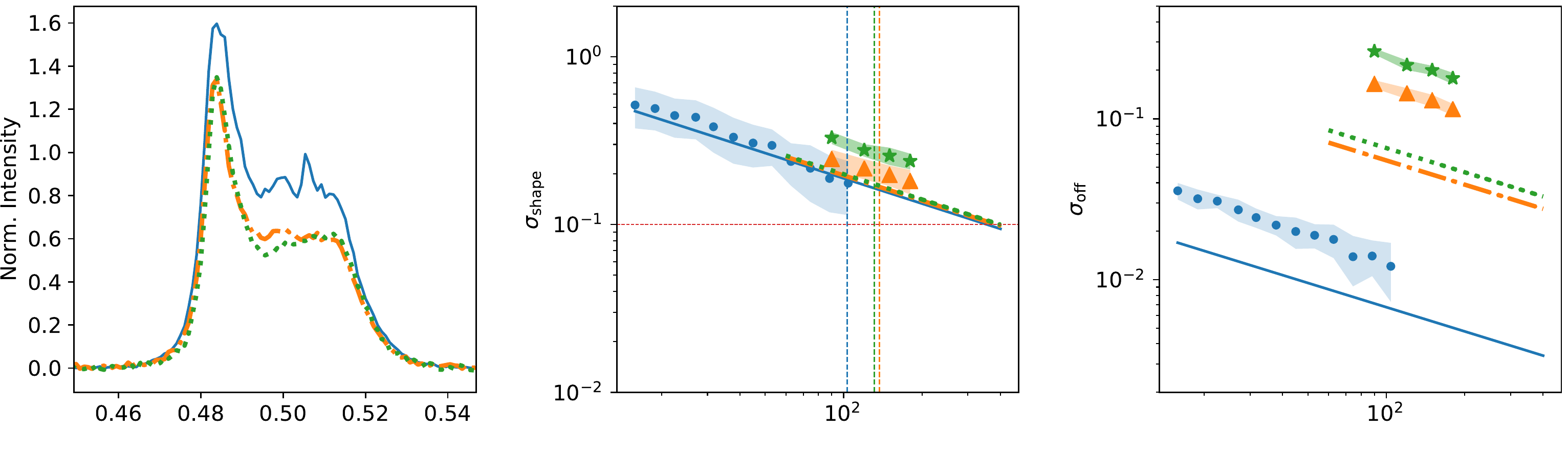}
	\includegraphics[width=\linewidth]{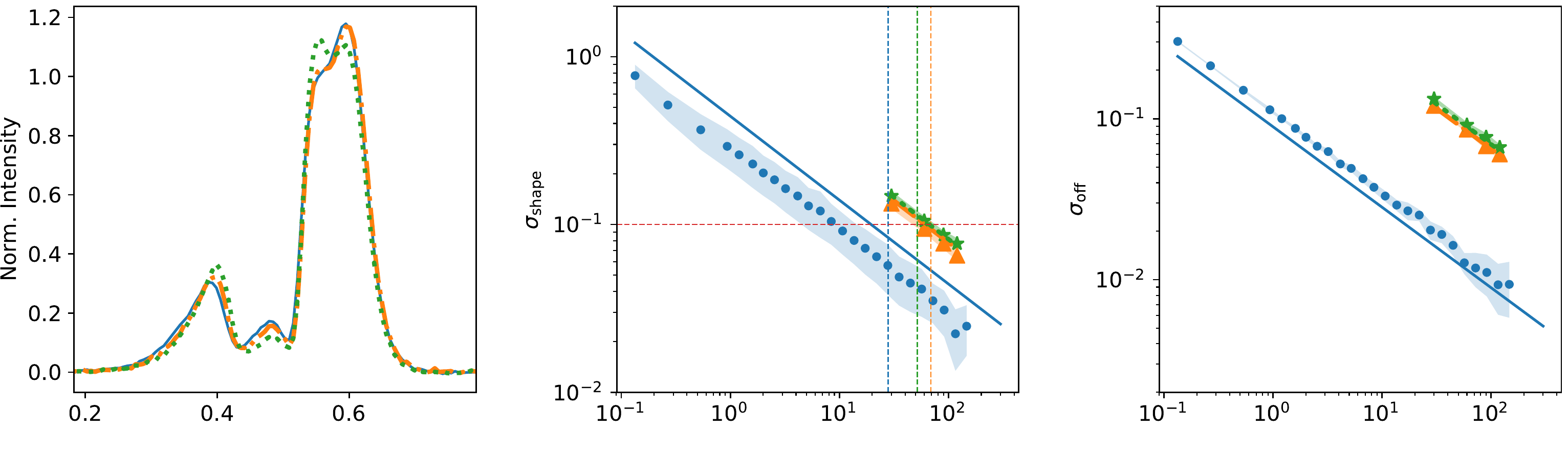}
	\includegraphics[width=\linewidth]{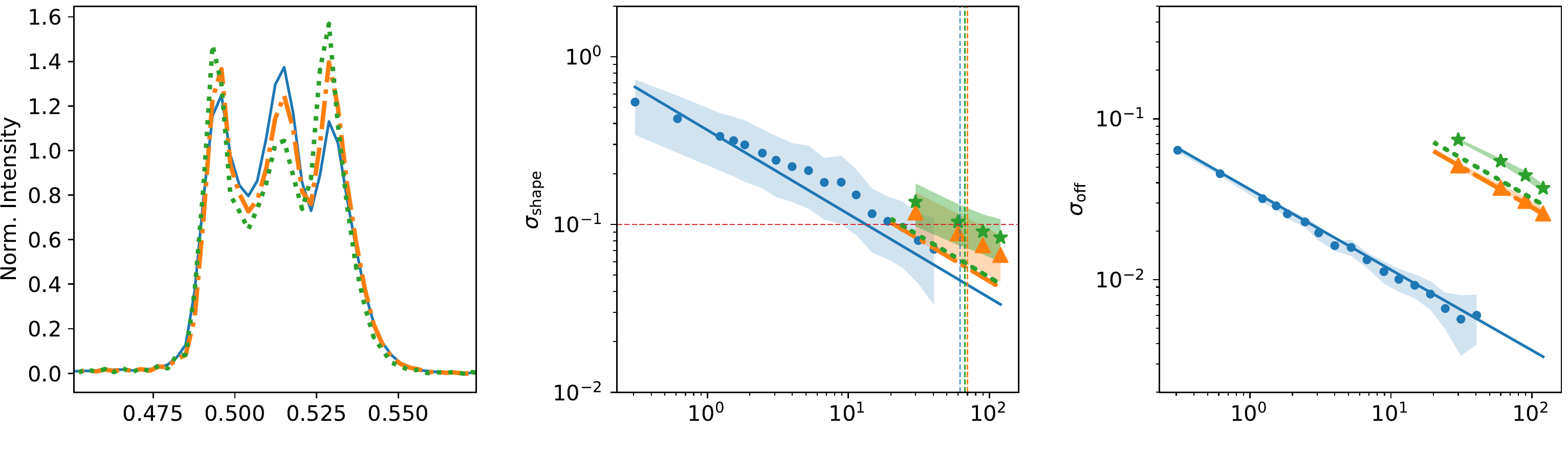}
	\includegraphics[width=\linewidth]{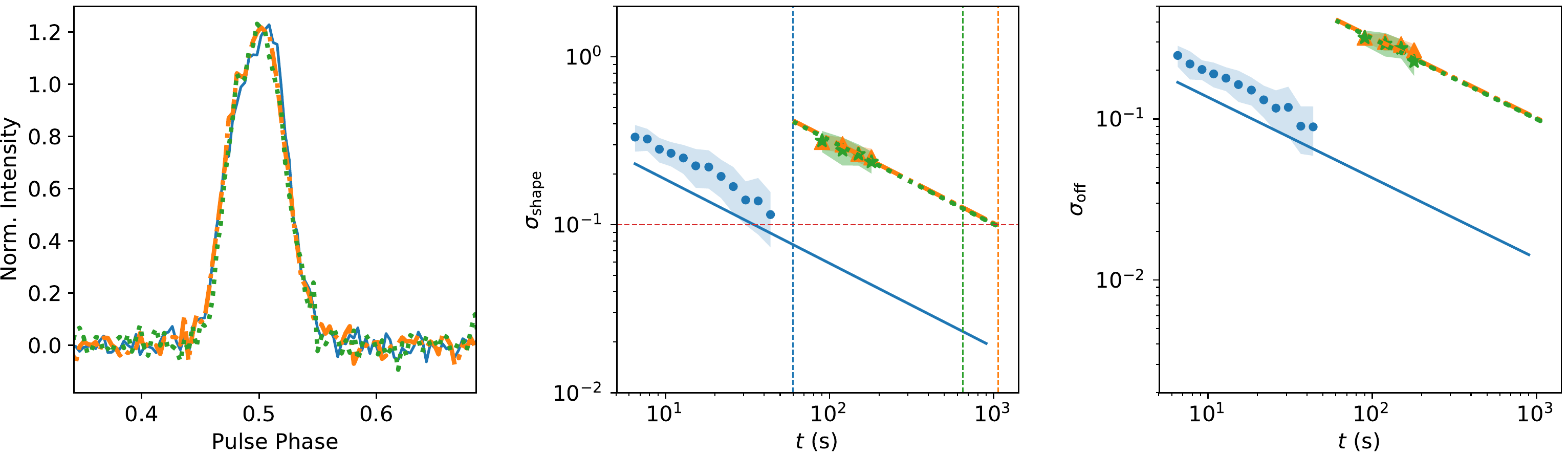}
	\caption{The measured $\sigma_{\mathrm{shape}}$ and $\sigma_{\mathrm{off}}$ for four pulsars (from top to bottom: PSRs J0729$-$1836, J1803$-$2137, J1829$-$1751 and J1841$-$0345) as a function of integration length. The left panel shows the normalised average pulse profile obtained from MeerKAT, Lovell and Parkes data in blue solid, green dotted and orange dash-dotted lines, respectively. The mean $\sigma_{\mathrm{shape}}$ (middle panel) and $\sigma_{\mathrm{off}}$ (right panel) obtained by measurements of 20 pairs of profile bins from MeerKAT, Lovell and Parkes data correspond to the blue dots, green stars and orange triangles, respectively. The standard deviations are shown as the shaded regions in the same colours. The diagonal lines (solid for MeerKAT, dotted for Lovell and dash-dotted for Parkes) show the predicted $\sigma_{\mathrm{shape}}$ using the weighted modulation index (see Table\,\ref{tab:para_psr}). For MeerKAT, the single-pulse mode bandwidth of 642 MHz was used. For Lovell and Parkes, the effective bandwidths of 200 MHz were assumed. The system temperature $T_{\mathrm{sys}}$ of the Lovell telescope was chosen to be the median value of all observations for a given pulsar by considering its elevation dependence. The vertical dashed lines with the same set of colours as the measurements mark $\tint$ (from left to right corresponding to MeerKAT, Lovell and Parkes telescopes). The horizontal red dashed line labels $\sigma_{\mathrm{shape}}=0.1$, the specified profile fidelity used to calculate $\tint$.}
	\label{fig:shaperms}
\end{figure*}

The results for PSR J0729$-$1836 are shown in the top panels of Fig.\,\ref{fig:shaperms}. The measured $\sigma_{\mathrm{shape}}$ (middle panel) covers an extended range as indicated by the shaded region, which corresponds to the standard deviation in $\sigma_{\mathrm{shape}}$ as determined for different pairs of on-pulse bins. This extend is the result of the details of the profile shape and the pulse longitude dependency of $m$. The correlated pulse-to-pulse variability between different profile bins plays a role as well. The dependence of $\sigma_{\mathrm{shape}}$ with integration times agrees with the prediction for the MeerKAT telescope (in blue solid line), while it is higher than the prediction for the Lovell and the Parkes data (in green dotted and orange dash-dotted lines respectively). This is because the measured system noise (top right panel) for all telescopes is about two times higher than assumed. This higher than expected system noise makes it comparable to the pulse-to-pulse variability in the Lovell and Parkes data, thus it contributes significantly to the measured $\sigma_{\mathrm{shape}}$ for those telescopes. For the more sensitive MeerKAT telescope, $\sigma_{\mathrm{shape}}$ remains dominated by the pulse-to-pulse variability. The fact that the measured system noise is systematically higher for all three telescopes suggests that the flux density $S_{1400}$ assumed is too high as obtained from the pulsar catalogue. This partially explains that why at $\tint$ (vertical lines), $\sigma_{\mathrm{shape}}$ is larger than the expected 0.1. In addition, $\tint$ is calculated using $m=1$ while $\bar{m}$ for this pulsar is about twice as large, making the pulse-to-pulse variability exceptionally large for this pulsar.

For PSR J1803$-$2137, as shown in the second row of panels in Fig.\,\ref{fig:shaperms}, the measured $\sigma_{\mathrm{shape}}$ and $\sigma_{\mathrm{off}}$ match well with the predicted trend for all three telescopes. The modulation index for this pulsar is 0.8, close to $m=1$ assumed in obtaining $\tint$. Therefore the $\sigma_{\mathrm{shape}}$ measured at $\tint$ is near the expected value of 0.1. 

In the case of PSR J1829$-$1751, presented in the third row of panels in Fig.\,\ref{fig:shaperms}, the observed trend of $\sigma_{\mathrm{shape}}$ with integration time follows the predicted trend at a level indicated by the shaded region for MeerKAT and Parkes. However, the Lovell measured $\sigma_{\mathrm{shape}}$ is too high, which is because $\sigma_{\mathrm{off}}$ is higher than predicted. Taking a more representative system temperature $T_{\mathrm{sys}}$ for these observations would make the measured $\sigma_{\mathrm{shape}}$ consistent with the prediction. Since $\bar{m} < 1$, $\tint$ can be expected to be conservative. Indeed, the measured $\sigma_{\mathrm{shape}}$ is 0.1 or lower at $\tint$. 

The bottom row panels show the result of PSR J1841$-$0345, a pulsar which is system noise dominated for the Lovell and Parkes telescopes. Although $\sigma_{\mathrm{shape}}$ and $\sigma_{\mathrm{off}}$ follow the $1/\sqrt{t}$ scaling, the system noise is larger than the prediction for MeerKAT. This affects the measured $\sigma_{\mathrm{shape}}$ accordingly. The higher measured $\sigma_{\mathrm{off}}$ could be due to the actual flux density of the pulsar for this MeerKAT observation being lower than the catalogue value (although there is no noticeable difference in the pulse profile shape), or the system temperature being higher. Taking a more representative $T_{\mathrm{sys}}$ would explain the offset between the measured $\sigma_{\mathrm{shape}}$ and the prediction. A $\sigma_{\mathrm{shape}} \simeq 0.1$ is achieved for all three telescopes at $\tint$, although this relies on extrapolation for the Lovell and Parkes telescopes. In the case of this MeerKAT observation, the higher than assumed system noise contribution is offset by a modulation index which is lower than 1.

So it is established that $\sigma_{\mathrm{shape}}$ follows an expected $1/\sqrt{t}$ scaling, and eq.\,\ref{eq:tint} can be applied in good approximation to pulsars with a wide range of parameters. As expected from the system properties, the full array of MeerKAT provides higher fidelity profiles compared to the Lovell and Parkes telescopes at a given integration time. For many pulsars $\tint$ will be conservative, although for pulsars with an exceptional high modulation index, e.g.\, PSR J0729$-$1836, this is not true. For pulsars for which the system noise dominates over pulse-to-pulse variability, $\tint$ will be more accurate since it is less affected by pulsar parameters. Nevertheless, the flux density remains a source of uncertainty. In addition, uncertainty in RFI conditions or variability in system performance is a source of uncertainty in the profile fidelity obtained. 

\subsection{Pulse widths measurements}
\label{app:width}
\begin{figure}
	\centering
	\includegraphics[width=\linewidth]{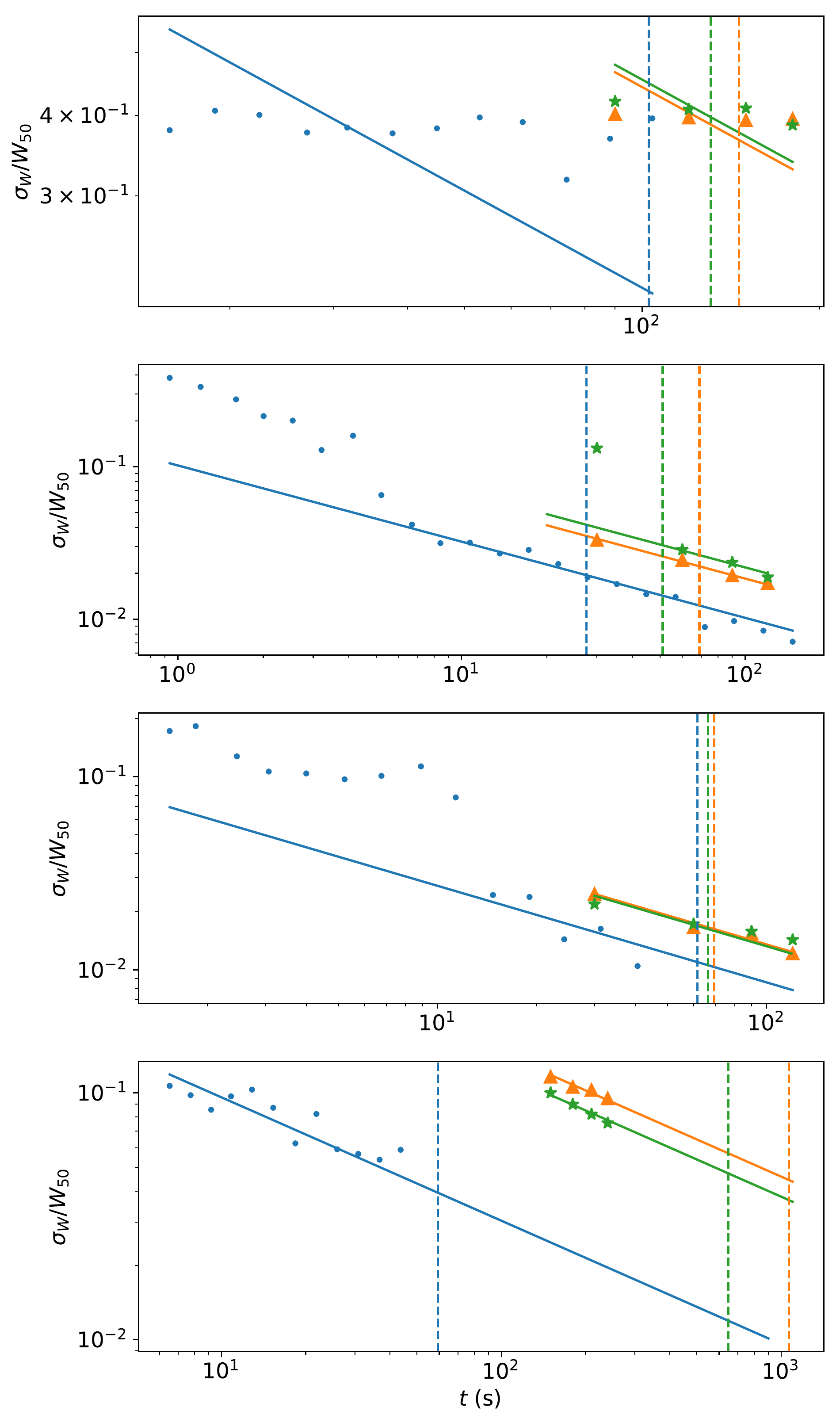}
	\caption{The measured fractional uncertainty of $W_{50}$ for the four pulsars (from top to bottom: PSRs J0729$-$1836, J1803$-$2137, J1829$-$1751 and J1841$-$0345) as a function of integration length. The results obtained from MeerKAT, Lovell and Parkes data are shown in blue dots, green stars and orange triangles respectively. The diagonal lines show a $1/\sqrt{t}$ dependence to guide the eye. The vertical dashed lines with the same set of colours as the measurements (from left to right corresponding to MeerKAT, Lovell and Parkes telescopes) mark $t=\tint$.}
	\label{fig:wthrms}
\end{figure}

The uncertainty of pulse width $W_{50}$, $\sigma_{W}$, is determined for the four pulsars at different integration lengths for MeerKAT, Lovell and Parkes observations. The $W_{50}$ measurements are done by the procedure explained in Section\,\ref{sec:w50}. Fig.\,\ref{fig:wthrms} shows the resulting $\sigma_{W}$ for the four pulsars, which for completeness includes the results for PSR J1803$-$2137 presented in Fig.\,\ref{fig:J1803_rmswth} as well. 

Fig.\,\ref{fig:wthrms} demonstrates that $W_{50}$ can be determined with a precision similar to $\Delta_{\mathrm{shape}}$ since $\sigma_{W}/W_{50} \lesssim 0.1$ for most pulsars at $\tint$ (marked as vertical dashed lines). This once more shows that $\tint$ is defined conservatively. A notable exception is PSR J0729$-$1836 (top panel of Fig.\,\ref{fig:wthrms}) for which $\sigma_{W}$ is about four times higher without any improvement with increasing integration length. This is because of the large pulse-to-pulse variability ($\bar{m}=1.9$) of this pulsar in combination with its particular profile shape (see top-left panel of Fig.\,\ref{fig:shaperms}), which has a trailing component with an amplitude about half of the main peak. As a consequence of profile variability, the trailing component is included often, but not always, in the $W_{50}$ measurement. The result is that the measured $W_{50}$ distribution is bi-modal with a large $\sigma_{W}$ which does not improve with integration length. This stresses the fact that $W_{50}$ is a complicated metric which depends on the actual pulse shape of the pulsar. In this particular case, the large measured $\sigma_{W}$ does not imply a low profile fidelity. 

For both PSRs J1803$-$2137 and J1829$-$1751, $\sigma_{W}$ deviates from the $1/\sqrt{t}$ scaling at short integration lengths as probed by the MeerKAT observations (blue dots in the second and third panel of Fig.\,\ref{fig:wthrms} respectively). The case of PSR J1803$-$2137 was discussed in Section\,\ref{sec:w50} and for PSR J1829$-$1751 the situation is similar. At these integration lengths the relative intensities of the three peaks in the pulse profile (see left panel in the third row of Fig.\,\ref{fig:shaperms}) become variable enough to affect the $W_{50}$ measurements in a similar way the PSR J0729$-$1836 measurements were affected at all probed integration lengths. After integrating long enough, the width is always determined by the three profile components and the $1/\sqrt{t}$ trend become evident. At shorter integration lengths, $W_{50}$ can measure the width of less than the three components combined, resulting in a high uncertainty. 

In general, as expected, given its larger gain, the measured $\sigma_{W}$ for MeerKAT is smaller than that for Lovell and Parkes at the same integration length. Unless the accuracy of $W_{50}$ is not a good measure for profile fidelity, its fractional uncertainty is comparable to that of $\Delta_{\mathrm{shape}}$, demonstrating that the methodology of determining $\tint$ is a robust way to obtain profiles at the desired fidelity. 

\subsection{Timing precision measurements}
\label{app:timingrms}

\begin{figure}
	\centering
	\includegraphics[width=\linewidth]{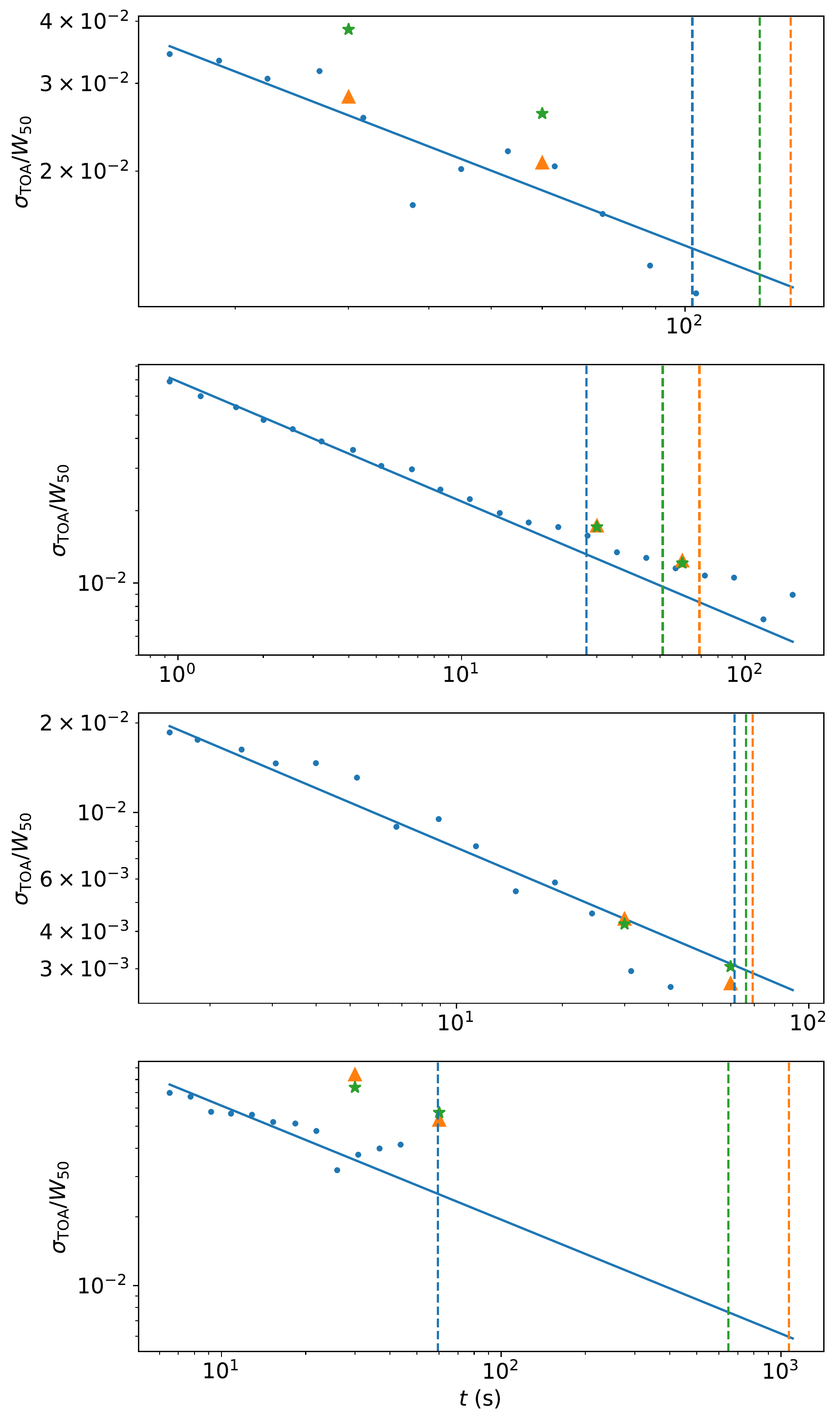}
	\caption{The achieved TOA precision of the four pulsars (from top to bottom: PSRs J0729$-$1836, J1803$-$2137, J1829$-$1751 and J1841$-$0345) as a function of integration time, normalised by $W_{50}$. See Fig.\,\ref{fig:wthrms} for an explanation of the different points and lines, which refer to the MeerKAT, Lovell and Parkes telescopes respectively.}
	\label{fig:resrms}
\end{figure}

Following the method outlined in Section\,\ref{sec:toa}, the timing precision $\sigma_{\mathrm{TOA}}$ is measured for different integration times for the three telescopes, based on the size of the timing residuals obtained. The results are shown in Fig.\,\ref{fig:resrms} and in all cases, a $1/\sqrt{t}$ dependence is observed. 

The main conclusion is that at $\tint$ (vertical lines) the fractional timing precision (relative to the pulse width $W_{50}$) is always very good at a level of a few per-cent. This is true for all pulsars as observed by all telescopes and exceeds the precision of the other shape parameters considered. As discussed in Section\,\ref{sec:toa}, this can be expected since TOAs are measured by comparing the pulse profile with an analytic template. This optimally uses the characteristics of the profile shape to form a single measurement, which therefore is less affected by the measurement which quantifies only one aspect of the pulse profile (such as a width). 

In Fig.\,\ref{fig:resrms}, for PSR J0729$-$1836 (top panel), the $\sigma_{\mathrm{TOA}}$ measurements are comparable for MeerKAT and Parkes, while they are higher for the Lovell data at the same timescale. The higher timing uncertainty for Lovell reflects that for this pulsar the system noise is relatively high (as can be seen in the $\sigma_{\mathrm{shape}}$ and $\sigma_{\mathrm{off}}$ measurements in the first row of Fig.\,\ref{fig:shaperms}), resulting in a lower S/N of a profile at a given integration time. For MeerKAT and Parkes, the pulse-to-pulse variability plays a larger role, resulting a timing precision which is comparable for the two telescopes despite the difference in sensitivity. As explained in Section\,\ref{sec:toa}, the same reason explains why for PSR J1803$-$2137 (second panel of Fig.\,\ref{fig:resrms}) the timing precision is comparable for all three telescopes, as profile variability dominates for each of them. Here it should also be noted, as explained in Section\,\ref{sec:toa}, that the TOA measurements are less affected by a system noise contribution than for other shape parameter measurements such as $\Delta_{\mathrm{shape}}$. From the third panel of Fig.\,\ref{fig:resrms}, one can see that also for PSR J1829$-$1751 $\sigma_{\mathrm{TOA}}$ is independent of the telescope used, and the reason is the same as for PSR J1803$-$2137.

On the other hand, PSR J1841$-$0345 is system noise dominated for the Lovell and Parkes telescopes, while this is not the case for MeerKAT. Therefore $\sigma_{\mathrm{TOA}}$ is expected to be higher for the Lovell and Parkes telescopes compared to MeerKAT, which is indeed seen in the fourth panel of Fig.\,\ref{fig:wthrms}.

\bsp	
\label{lastpage}
\end{document}